\documentstyle[preprint,prl,aps]{revtex}

\begin{document}
 
\preprint{SLAC-PUB-7994, July 2000 T/E}
 
\title{Measurements of the $Q^2$-Dependence of the Proton and Neutron Spin
Structure Functions  $ g_1^p$ and $ g_1^n$}
 
\author{The E155 Collaboration:
P.~L.~Anthony,$^{16}$  
R.~G.~Arnold,$^{1,12}$
T.~Averett,$^{5,\diamond}$
H.~R.~Band,$^{21}$
M.~C.~Berisso,$^{12}$
H.~Borel,$^7$
P.~E.~Bosted,$^{1,12}$
S.~L.~B${\ddot {\rm u}}$ltmann,$^{19}$
M.~Buenerd,$^{16,\dag}$
T.~Chupp,$^{13}$
S.~Churchwell,$^{12,\ddag}$
G.~R.~Court,$^{10}$
D.~Crabb,$^{19}$
D.~Day,$^{19}$
P.~Decowski,$^{15}$
P.~DePietro,$^1$
R.~Erbacher,$^{16,17}$
R.~Erickson,$^{16}$
A.~Feltham,$^{19}$
H.~Fonvieille,$^3$
E.~Frlez,$^{19}$
R.~Gearhart,$^{16}$
V.~Ghazikhanian,$^6$
J.~Gomez,$^{18}$
K.~A.~Griffioen,$^{20}$
C.~Harris,$^{19}$
M.~A. Houlden,$^{10}$
E.~W.~Hughes,$^5$
C.~E.~Hyde-Wright,$^{14}$
G.~Igo,$^6$
S.~Incerti,$^3$
J.~Jensen,$^5$
J.~R.~Johnson,$^{21}$
P.~M.~King,$^{20}$
Yu.~G.~Kolomensky,$^{5,12}$
S.~E.~Kuhn,$^{14}$
R.~Lindgren,$^{19}$
R.~M.~Lombard-Nelsen,$^7$
J.~Marroncle,$^7$
J.~McCarthy,$^{19}$
P.~McKee,$^{19}$
W.~Meyer,$^{4}$
G.~S.~Mitchell,$^{21,\times}$
J.~Mitchell,$^{18}$
M.~Olson,$^{9,\Box}$
S.~Penttila,$^{11}$
G.~A.~Peterson,$^{12}$
G.~G.~Petratos,$^9$
R.~Pitthan,$^{16}$
D.~Pocanic,$^{19}$
R.~Prepost,$^{21}$
C.~Prescott,$^{16}$
L.~M.~Qin,$^{14}$
B.~A.~Raue,$^{8,18}$
D.~Reyna,$^{1,\circ}$
L.~S.~Rochester,$^{16}$
S.~Rock,$^{1,12}$
O.~A.~Rondon-Aramayo,$^{19}$
F.~Sabatie,$^7$
I.~Sick,$^2$
T.~Smith,$^{13,\times}$
L.~Sorrell,$^1$
F.~Staley,$^7$
S.~St.Lorant,$^{16}$
L.~M.~Stuart,$^{16,\S}$
Z.~Szalata,$^1$
Y.~Terrien,$^7$
A.~Tobias,$^{19}$
L.~Todor,$^{14}$
T.~Toole,$^1$
S.~Trentalange,$^{6}$
D.~Walz,$^{16}$
R.~C.~Welsh,$^{13}$
F.~R.~Wesselmann,$^{14,19}$
T.~R.~Wright,$^{21}$
C.~C.~Young,$^{16}$
M.~Zeier,$^2$
H.~Zhu,$^{19}$
B.~Zihlmann,$^{19}$
}
\address{
{$^{1}$American University, Washington, D.C. 20016}  \break
{$^{2}$Institut f${\ddot u}$r Physik der Universit${\ddot a}$t Basel, CH-4056 Basel, Switzerland} \break
{$^{3}$University Blaise Pascal, LPC IN2P3/CNRS F-63170 Aubiere Cedex, France} \break
{$^{4}$Ruhr-Universit${\ddot a}$t Bochum, Universit${\ddot a}$tstr. 150, Bochum, Germany} \break
{$^{5}$California Institute of Technology, Pasadena, California 91125}\break
{$^{6}$University of California, Los Angeles, California 90095} \break
{$^{7}$DAPNIA-Service de Physique Nucleaire, CEA-Saclay, F-91191 Gif sur Yvette, France} \break
{$^{8}$Florida International University, Miami, Florida 33199} \break
{$^{9}$Kent State University, Kent, Ohio 44242} \break
{$^{10}$University of Liverpool, Liverpool L69 3BX, United Kingdom } \break
{$^{11}$Los Alamos National Laboratory, Los Alamos, New Mexico 87545} \break
{$^{12}$University of Massachusetts, Amherst, Massachusetts 01003} \break
{$^{13}$University of Michigan, Ann Arbor, Michigan 48109} \break
{$^{14}$Old Dominion University, Norfolk, Virginia 23529} \break
{$^{15}$Smith College, Northampton, Massachusetts 01063} \break
{$^{16}$Stanford Linear Accelerator Center, Stanford, California 94309 } \break
{$^{17}$Stanford University, Stanford, California 94305} \break
{$^{18}$Thomas Jefferson National Accelerator Facility, Newport News, Virginia 23606} \break
{$^{19}$University of Virginia, Charlottesville, Virginia 22901} \break
{$^{20}$The College of William and Mary , Williamsburg, Virginia 23187} \break
{$^{21}$University of Wisconsin, Madison, Wisconsin 53706} \break
}
 
\maketitle

\begin{abstract}
   The structure functions $g_1^p$ and $g_1^n$ have 
been measured over the range
$0.014<x<0.9$ and $1<Q^2<40$ GeV$^2$ using deep-inelastic
scattering of 48 GeV longitudinally polarized electrons from 
polarized protons and deuterons. We find that the
 $Q^2$ dependence of $g_1^p$ 
($g_1^n)$ at fixed $x$ is very similar to that of the spin-averaged
structure function  $F_1^p$ ($F_1^n$ ).
From an NLO QCD fit to all available data we find 
 $\Gamma_1^p - \Gamma_1^n =0.176\pm0.003\pm 0.007$ 
at $Q^2=5$~GeV$^2$,
in agreement with the Bjorken sum rule prediction of $0.182\pm0.005$.
   \end{abstract}
 
 
\maketitle 

\narrowtext
 
     The spin-dependent
structure function $g_1(x,Q^2)$  for deep-inelastic
lepton-nucleon scattering is of fundamental importance  in understanding  the
quark and gluon spin structure of the proton and neutron. The $g_1$ structure
function depends both on $x$, the fractional momentum carried by the struck
parton, and on $Q^2$, the squared four-momentum  of 
the exchanged virtual photon.
The fixed-$Q^2$ integrals (or first moments) 
$\Gamma^p_1(Q^2)=\int_0^1g_1^p(x,Q^2)dx$ for the proton and
$\Gamma^n_1(Q^2)=\int_0^1g_1^n(x,Q^2)dx$ for the neutron
are related to the net quark helicity $\Delta \Sigma$ in the
nucleon. Measurements
of $\Gamma^p_1$ \cite{EMC,SMC,E143,HERMES},  
$\Gamma^d_1$ for the deuteron \cite{SMC,E143,E155} 
(which essentially measures the average of the proton
and neutron), and
$\Gamma^n_1$ \cite{HERMES,E142,E154,E154fit} have 
found $\Delta \Sigma$ between 0.2 and 0.3,  
significantly less than the prediction \cite{ellisjaffe} 
that $\Delta \Sigma=0.58$ assuming zero net  strange
quark helicity and  SU(3) flavor symmetry in the baryon
octet.  A fundamental sum
rule originally derived  from current algebra by 
Bjorken \cite{bjorken} predicts
 $\Gamma^p_1(Q^2)-\Gamma^n_1(Q^2)=g_A/6g_V$. 
Recent measurements are in
agreement with this sum rule prediction when 
perturbative QCD (pQCD) corrections
\cite{larin} are included.
 
  According to the DGLAP equations
\cite{GLAP}, $g_1$  is expected to evolve logarithmically with $Q^2$,
and in the case of $g_1^p$ to 
increase with $Q^2$ at low $x$, and  decrease with $Q^2$ at
high $x$ \cite{E143}. A similar $Q^2$-dependence has been observed in
 the spin-averaged
structure functions $F_1(x,Q^2)$, while the ratio $g_1/F_1$ has been found
to be  approximately independent of $Q^2$ \cite{E143}.
The precise behavior is sensitive to the  underlying spin-dependent quark
and gluon distribution functions. Fits to  data for $g_1$ using NLO pQCD 
allow  determinations of the first moments (from which the
Bjorken sum rule can be tested) as well as the valence quark, sea quark, 
and gluon spin contributions. The goal of the present experiment
(SLAC E155) was to make
precise measurements over a wide range of $Q^2$ in a single experiment
to further constrain these quantities. 

 The ratio of polarized to unpolarized structure functions
can be  determined from measured longitudinal asymmetries $A_\|$ using
\begin{equation} 
      g_1/F_1= A_{\|}/d+(g_2/F_1)[(2Mx)/(2E-\nu)],
\end{equation}
where $d=[(1-\epsilon)(2-y)]/ \{y[1+\epsilon R(x,Q^2)]\}$,  $y=\nu/E$,
and $\nu=E-E^\prime$, where $E$ is the incident and 
$E^\prime$ is the scattered electron energy in the lab frame,
$\epsilon^{-1}=1+2[1+\gamma^{-2}]\tan^2(\theta/2)$, $\gamma^2=Q^2/\nu^2$,
$\theta$ is the electron scattering angle,
$M$ is the nucleon mass, and
$R(x,Q^2)=[F_2(x,Q^2)(1+\gamma^2)]/[2xF_1(x,Q^2)]-1$ is typically 0.2 for the
kinematics of this experiment \cite{whitlowr}.  
For the contribution of the transverse
spin structure function $g_2$ we used the twist-two result of Wandzura
and  Wilczeck ($g_2^{WW}$) \cite{WW}
\begin{equation} g_2^{WW}(x,Q^2)=-g_1(x,Q^2)+\int_x^1 g_1(\xi,Q^2)d\xi/\xi,
\end{equation} evaluated using the empirical  fit to $g_1/F_1$ given below
(Eqs. 5 and 6).
The  $g_2^{WW}$ model is in good agreement with existing data
\cite{E143,E155g2,E155x}. Using other reasonable models for $g_2$ that
agree with existing data makes negligible changes to the
extracted $g_1/F_1$ values due to suppression of the $g_2$ contribution
by the factor $2Mx/(2E-\nu)$.
The $g_1$ and $g_2$
structure functions are related to 
 the virtual photon asymmetry $A_1=(g_1/F_1)-\gamma^2(g_2/F_1)$ (which is
bounded by $\vert A_1 \vert \leq 1$).
 
In this Letter we report new measurements of  $g_1^p$ made using a
48.35 GeV polarized electron beam at SLAC. 
The new data extend to higher
$Q^2$ (40 GeV$^2$) and lower $x$ (0.014) than previous high statistics
SLAC measurements \cite{E143}. Combined with measurements of 
$g_1^d$ made in this same  experiment using a $^6$LiD target \cite{E155}, 
we can extract $g_1^n$ and compare 
with E154 \cite{E154} which measured  $g_1^n$  at 
similar kinematics using a polarized $^3$He target as a source of polarized
neutrons. 

Longitudinally polarized electrons were produced by
photoemission from a strained-lattice GaAs  crystal.  
Beam pulses were typically 0.3 $\mu$s long,
contained 2--4$\times 10^{9}$ electrons, and were delivered at a
rate of 120 Hz. The helicity was selected randomly on a
pulse-to-pulse basis to minimize instrumental
asymmetries. The longitudinal beam polarization $P_b$ was measured
using M\o ller scattering from thin, magnetized ferromagnetic foils,
periodically inserted about 25 m before the polarized  target used
to measure $g_1$. 
Results from two detectors (one detecting a single
final-state electron, 
the other detecting two electrons in  coincidence)
agreed within errors, yielding $P_b=0.813\pm0.020$.
 
As in E143 \cite{E143}, the 3-cm-long polarized target  cell  
contained  pre-irradiated granules of 
$^{15}$NH$_3$ immersed in liquid He at  1 K in  a uniform 
magnetic field of 5 T.
Microwaves near 140
GHz were used to drive the hyperfine transition which 
aligns (or anti-aligns) the
nucleon spins with the magnetic field, producing  proton
polarizations of typically  90\%  in 10 to 20 minutes.  The
polarization slowly decreased due to radiation damage, and was
periodically restored by  annealing the target at
about 80 K\null. The  2-3 mm diameter electron beam spot was
rastered over the 3 cm$^2$ front
surface of the target to uniformly distribute
beam heating and radiation damage.
To study possible experimental biases,  the target polarization 
direction was periodically reversed using slight
 adjustments to  the microwave frequency.
Also, the direction of the magnetic field was reversed several
times during the experiment. Final asymmetry results  were consistent 
for the four polarization/field direction
combinations.
 
The target polarization $P_t$ was monitored with the same NMR Q-meter
system as was used in experiment E143 \cite{E143}.  The E143 design 
of target cell was modified for this experiment to improve the target 
polarization  (average value of $P_t$ was about 0.8 for E155) 
and this change had unforeseen effects on the
performance of the NMR system when it was used to measure the proton
polarization. Consequently, the NMR system was operated outside its 
design envelope, resulting in a significant degree of non-linear
behavior. This problem is now qualitatively understood \cite{court} but 
insufficient information about the NMR RF circuit parameters is
 available to allow adequate corrections for these non-linear effects
to be calculated. Therefore the polarization data, for the proton target
only, was extracted using the observed dependence of the polarization on the
integrated beam dose deposited in the target material 
% \cite{crday}
obtained primarily from experiment E155x \cite{E155x}. This method leads to 
a larger systematic error in the proton polarization 
measurements (typically 7\%) than would have been
obtained using the standard NMR technique. It should be emphasized that
this problem is unique to this particular set of proton experimental 
data and was eliminated in experiment E155x \cite{E155x}  by a further 
target cell design change.

Scattered electrons
were detected in 
three independent magnetic spectrometers centered at angles of 2.75,
5.5, and 10.5 degrees.
The two small angle spectrometers
were the same as in E154 \cite{E154}, while the large angle spectrometer
was new for this experiment. It was composed of a single dipole magnet
and two quadrupoles, and covered $7<E^\prime <20$ GeV, 
$9.6^\circ < \theta<12.5^\circ$,
and $-18<\phi<18$ mr, for a maximum solid angle of 1.5 msr at 8 GeV. Electrons
were separated from a much larger flux of pions by using a gas Cherenkov counter
and a segmented lead glass electromagnetic calorimeter.  

 The experimental asymmetries  $A_{\|}$   were determined from
\begin{equation}
A_\| =\bigg({N_- -N_+ \over
N_- +N_+} \bigg) {C_{N} \over f P_b P_t f_{RC}}+A_{RC},
\end{equation}
where the target polarization is parallel
to the beam direction, $N_-$ 
($N_+$) is the number of scattered electrons per incident charge for negative
(positive) beam helicity,  $C_N \approx 0.985$ is a 
correction factor for the
polarized nitrogen nuclei, $f$ is the dilution factor representing the fraction of
measured events originating from polarizeable hydrogen within the
target, and $f_{RC}$ and $A_{RC}$ take into account  radiative corrections.
 
The dilution factor $f$ varied with $x$ between 0.13 and 0.17; it was
determined from a detailed model of
 the number of measured counts expected from
each component of the  target, including $^{15}$NH$_3$, various windows, 
NMR coils, liquid helium, etc. A typical target  contained about 13\% free
protons, 66\% $^{15}$N, 10\% $^4$He, 6\% Al, and 5\% Cu-Ni
by weight.  The relative systematic
error in $f$ ranges from 2.2\% to 2.6\%.
 
A correction to the asymmetries was made for hadrons misidentified as
electrons (typically 2\% of electron candidates, but up to 15\% in
the lowest $x$ bin of the 10.5$^\circ$ spectrometer). The correction used
the asymmetry measured for a large sample of inclusive hadrons, which
was found to be close
to zero at all kinematics. An additional correction was made for 
electrons from pair-symmetric processes
(such as $e^+/e^-$
pair production from photons) measured by reversing the spectrometer
polarity. The measured  pair-symmetric $A_\|$ 
was consistent with zero at all kinematics,
so the correction is equivalent to a dilution factor correction of
typically 10\% at the lowest $E^\prime$ of each spectrometer, decreasing
rapidly to a negligible correction at higher $E^\prime$.

Corrections were applied for the rate-dependence of the detector response,
which changed the measured asymmetries by less than 1\%. Corrections for
kinematic resolution were generally a few percent or less, except 
for $x>0.6$ where corrections to the measured
asymmetries  were as large as 15\%.  
 
The internal radiative corrections for $A_\|$ 
were  evaluated using the formulae of
Kuchto and Shumeiko \cite{radcor}.
The cross sections entering the asymmetry were `externally 
radiated'  according to Tsai \cite{tsai}. Comparison of Born and 
fully radiated asymmetries allowed us to
extract the asymmetry corrections $f_{RC}$ and $A_{RC}$. 
By splitting 
the radiative correction into these two parts, we can propagate 
consistently the experimental error to the extracted Born asymmetries 
for the corresponding kinematic bins, in the
presence of `dilution' from elastic and inelastic radiative 
tails. Previous
analyses (including E143) have used values of $f_{RC}$
 closer to unity by taking only 
quasi-elastic radiative tails into account, leading to 
smaller error bars at low $x$. Our new treatment is based on a 
definition of $f_{RC}$ that insures that the additive 
correction $A_{RC}$ is statistically
independent from the data point to which it is applied. 
Values for $f_{RC}$ range
from 0.45 at the lowest $x$-bin to greater 
than 0.9 for $x>0.15$, similar to the
results in E154. However, the resulting net
correction of the measured asymmetries is relatively small (0.01 to 0.02).
The E155 radiative corrections are based on an iterative 
global fit to all available data, in which all previous SLAC data were
re-corrected in a self-consistent way.

The E155 results for $g_1^p/F_1^p$ and $g_1^n/F_1^n$
are shown in Figs.~1 and 2  as a function
of $Q^2$ at eleven  values of $x$, 
and are listed in Table I. 
The neutron results were obtained from the
proton results and E155 deuteron results \cite{E155} using
\begin{equation}
 {g_1^n } = {g_1^d  \over 1-1.5\omega_D} { F_1^n + F_1^p  \over  F_1^d} -
g_1^p
\end{equation}
For the deuteron D-state probability we use $\omega_D=0.05\pm 0.01$,
and $ F_1^p / F_1^n$ was obtained from the NMC fit \cite{NMC}.
Slight changes  to the data of Refs. \cite{EMC,SMC} 
were made  to use the $g_2^{WW}$ model \cite{WW}  
for $g_2$ instead of assuming $A_2=0$. Data from all experiments 
\cite{EMC,SMC,E143,HERMES} have
been matched to the $x$ bins in Figs. 1 and 2 using the simple fit below for
small bin centering corrections.

 For the present experiment, most systematic errors 
(beam polarization, target polarization,
fraction of polarizeable nucleons in the target) for a given target are
common to all data  and correspond to an overall
normalization error of about 7.6\% for the proton data. 
The remaining systematic errors (model dependence of radiative corrections, 
model uncertainties for
$R(x,Q^2)$, resolution corrections) vary smoothly with $x$ 
in a locally correlated
fashion, ranging from a few percent for mid-range  $x$ bins, 
up to 15\% for the highest and lowest bins. 

%The theoretical interpretation of $g_1$ at low $Q^2$ is 
%complicated by higher
%twist contributions \cite{twist} not embodied in the 
%DGLAP equations. These terms are
%expected to be proportional to $C(x)/Q^2$, $D(x)/Q^4$, etc., 
%where $C(x)$ and
%$D(x)$ are unknown functions.  Analysis of previous data \cite{E143}
%suggests that these terms are small for $Q^2>1$ GeV$^2$ at moderate
%$x$ ($0.05<x<0.3$). We therefore have put a cut $Q^2>1$ GeV$^2$ 
%for the present  analysis (which only removes data from E143 and SMC, since
%all the data for E155 already pass this cut). 
 
Given the relatively large overall normalization uncertainty, 
the E155 data are in good agreement with the average
of world data \cite{EMC,SMC,E143,HERMES,E142,E154}. 
If we were to allow  an overall normalization factor for our proton
data, we would find a value of $1.08\pm 0.03$(stat)$\pm 0.07$(syst). 

 In any given $x$ bin, there is no evidence of strong
$Q^2$ dependence for the ratio $g_1/F_1$. A simple parametric  
fit to world data with $Q^2>1$ GeV$^2$ (to mitigate possible
higher twist contributions \cite{twist}) and missing mass $W>2$ GeV 
(to avoid complications from the resonance region),
shown as the dashed curves in all three figures, is given by 
\begin{equation}
{g_1^p \over F_1^p}=x^{0.700}(0.817+1.014x-1.489x^2)(1-{0.04\over Q^2})
\end{equation}
\begin{equation}
{g_1^n  \over  F_1^n}=x^{-0.335}(-0.013 -0.330x+0.761x^2)(1+{0.13\over Q^2}).
\end{equation}
This fit has an acceptable  $\chi^2$ of 478  for 483  degrees of freedom.
The coefficients of $-0.04\pm0.06$ ($0.13\pm0.45$) 
for the overall proton (neutron)
$Q^2$ dependence are small and consistent with zero. 

To examine the $x$ dependence of $g_1$ at fixed $Q^2$, we  averaged
the E155 results over $Q^2$ assuming the $Q^2$ dependence of the fit above
and use $F_1$ from \cite{whitlowr,NMC} to 
obtain results for $g_1^p$ and  $g_1^n$ at a fixed $Q^2=5$ GeV$^2$,
shown in Fig. 3. The proton data suggest $g_1$ is approximately
constant or slightly rising as $x\rightarrow 0$, but 
the neutron data are consistent with  the
trend of the E154 data to become increasingly negative at low $x$. 
The difference $g_1^p-g_1^n$ (which enters into the Bjorken sum rule)
is theoretically expected
to be well-behaved as $x \rightarrow 0$ compared to either 
$g_1^p$ or $g_1^n$. This is because if isospin is a good symmetry, the
sea quark and gluon contributions cancel, leaving only the
difference of $u$ and $d$ quark valence distributions. 
The errors on the present data are too large to clearly
support or contradict this expectation (see Fig. 3c). 

The choice of low$-x$ extrapolation has a large impact on the 
the evaluation of the first moment of $g_1$. 
To be consistent with other analyses of $g_1$,  
we have made a NLO pQCD fit in the $\overline{MS}$ scheme 
to all recent data, using assumptions similar to those in  \cite{E154fit}. 
The polarized parton distributions were parameterized as 
\begin{equation}
$$
\Delta f(x,Q_{0}^{2}) = A_{f}x^{{\alpha}_{f}} f(x,Q_{0}^{2}),
$$
\end{equation}
where $\Delta f = \Delta u_{v}$, $\Delta d_{v}$, $\Delta \overline{Q}$, 
and $\Delta G$ are the polarized valence, sea, and gluon distributions, 
and the $f(x,Q_{0}^{2})$ are the unpolarized parton distributions 
at $Q_0^2 = 0.40$ GeV$^2$ from
Ref. \cite{NEWGRV}. The positivity  constraint $\left| \Delta f \right|<f$ 
was imposed in this fit, as well as the requirement $\alpha_{f}>0$. 
The sea quark distributions were parameterized as 
$\Delta\overline{Q}={1 \over 2}(\Delta\overline{u}+
\Delta\overline{d})+{1 \over 5} \Delta \overline{s}$.
We assumed a symmetric quark sea for this analysis. 
We have not fixed the normalization of the
non-singlet distributions, so that the fit results 
test the Bjorken sum rule. However, $\alpha_s(M_Z^2)$ has been fixed at 0.114
for consistency with the unpolarized distributions that were used \cite{NEWGRV}. 
The fit results are: $A_u=0.95$, $A_d=-0.42$, 
$A_Q=0.01$, $A_g=0.50$, 
$\alpha_u=0.57$, $\alpha_d=0.0$, $\alpha_Q=1.00$, and $\alpha_g=0.02$. 
The overall $\chi^2$/d.f. is $1.10$ using statistical errors only. 
Evaluations of the fit are plotted as the solid curves
in Figs. 1-3, and indicate only
a slight dependence on $Q^2$ for $g_1/F_1$ in the $x$ region where there
are high statistics data. For $x<0.014$, the proton and neutron 
fits become increasingly 
negative at fixed $Q^2$ (see Fig. 3), 
although the difference stays closer to zero and makes only a small
contribution to the Bjorken sum rule. 

Using the NLO pQCD fit, we find the quark singlet 
contribution  $\Delta \Sigma=0.23\pm 0.04$(stat)$\pm 0.06 $(syst) 
at $Q^2=5$ GeV$^2$, 
well below the Ellis-Jaffe prediction \cite{ellisjaffe} of 0.58.
We find $\Gamma_1^p=0.118\pm0.004\pm 0.007$,
$\Gamma_1^n=-0.058\pm0.005\pm 0.008$, and 
 $\Gamma_1^p - \Gamma_1^n =0.176\pm0.003\pm 0.007$,  
in good agreement with the Bjorken sum rule 
prediction of $0.182\pm0.005$
evaluated with up to third order corrections in $\alpha_s$
\cite{larin}. For the first moment of the gluon
distribution we obtain $\Delta G=1.6 \pm 0.8 \pm 1.1$. The error
on this quantity is too large to significantly constrain the gluon 
contribution to the nucleon spin sum rule.

In summary, the new data on $g^p_1$ and $g^n_1$ extend the range of high
statistics electron scattering results to lower $x$ and 
higher $Q^2$ than previous data,
improving the errors obtained from NLO pQCD fits to world data. 
The Bjorken sum rule prediction is validated within errors, while the
extracted quark singlet contribution is small at approximately 0.2. 

This work was supported by the Department of Energy
(TJNAF, FIU, Massachusetts, ODU, SLAC, Stanford, Virginia, Wisconsin,
and William and Mary);  by the National Science
Foundation (American, Kent, Michigan, and ODU); by the
Schweizersche Nationalfonds (Basel); 
by the Kent State University Research Council (GGP); 
by the Commonwealth of Virginia (Virginia);  by
the Centre National de la Recherche Scientifique and the Commissariat a l'Energie
Atomique (French groups).

\begin{figure}
\vspace*{6.2in}
\hspace*{.45in}
\includegraphics{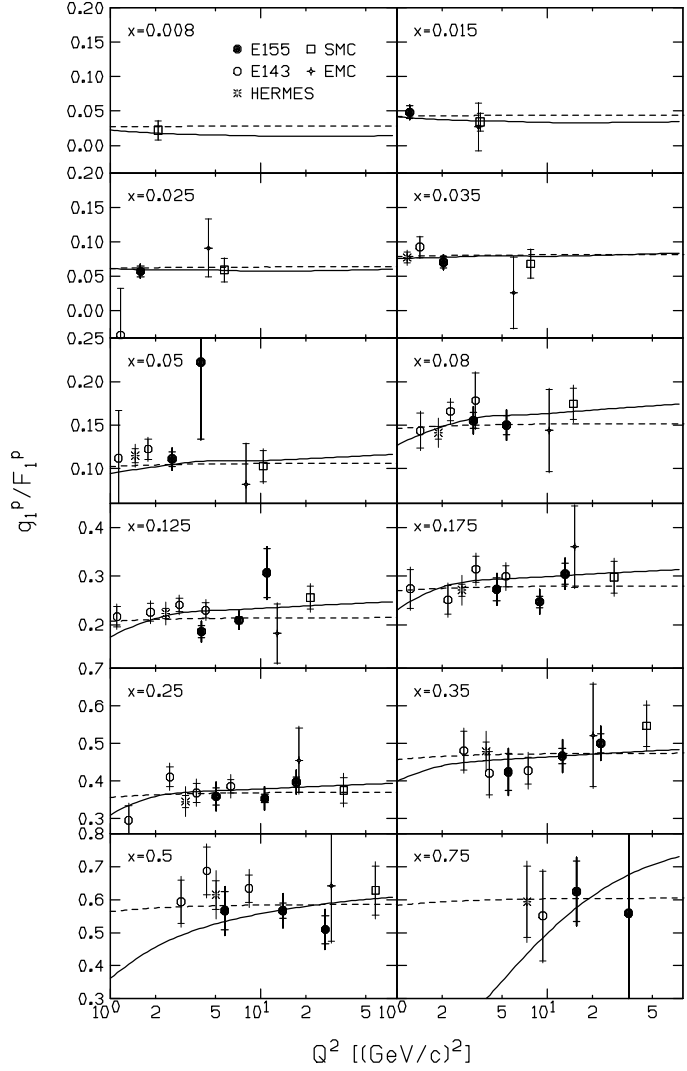}
\caption{Ratios $g_1^p/F_1^p$ extracted from 
experiments assuming the $g_2^{WW}$
model for $g_2$. Inner  errors are statistical only, while systematic errors
are included in quadrature in the outer error bars. 
The  solid curves correspond to the NLO QCD
fit described in the text, 
while the dashed curves are from the simple fit given by Eq. 5.}
% Data are from this experiment
%(solid circles),  HERMES \protect\cite{HERMES} (diamonds),
%E143 \protect\cite{E43} (triangles), 
%EMC \protect\cite{EMC} (squares), and SMC
%\protect\cite{SMC} (open circles). 
\end{figure}

\vfill\eject
\begin{figure}
\vspace*{6.2in}
\hspace*{.45in}
\includegraphics{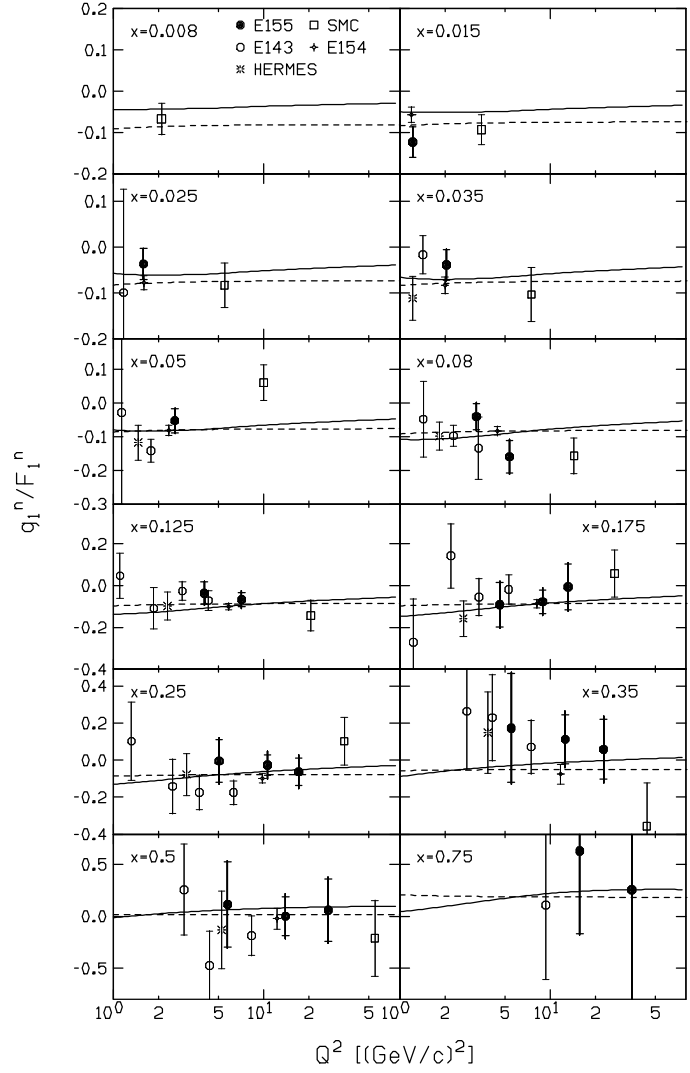}
\caption{Same as Fig. 1 except for $g_1^n/F_1^n$ and Eq. 6 for the
dashed curves.}
\end{figure}

\vfill\eject
\begin{figure}
\vspace*{6.2in}
\hspace*{.45in}
\includegraphics{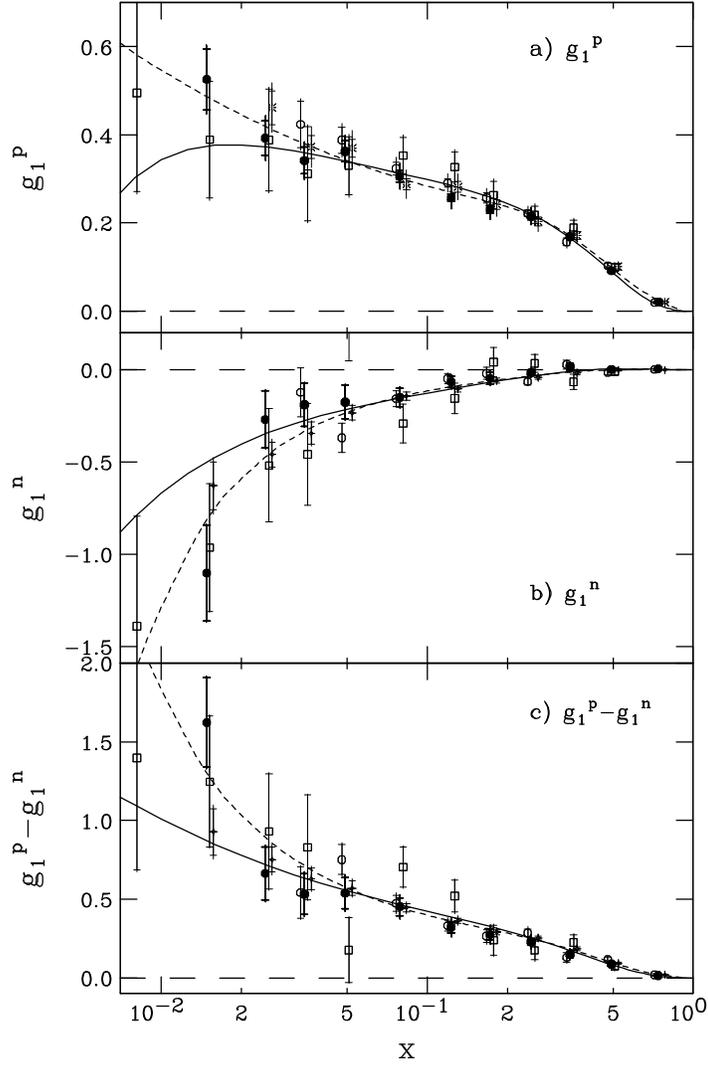}
\caption{ Data for $g_1^p$ (a), $g_1^n$ (b), and 
$g_1^p - g_1^n$ (c), 
evaluated at $Q^2=5$ GeV$^2$. 
The data are from this experiment (solid circles), E143 (open circles),
SMC (squares), HERMES (stars), and E154 (crosses). The 
$g_1^p - g_1^n$ values were obtained from the proton and deuteron results of
E155, E143, and SMC, while the proton E155 and neutron E154 results 
were used to obtain the results with the cross symbol.
The curves are as in Figs. 1 and 2.}
\end{figure}

\vfill\eject
\begin{table}[t]
\caption{Results for $g_1/F_1$ from this experiment for the proton
and neutron for $E=48.35$ and the indicated values of $x$ and $Q^2$.}
\begin{tabular}{cccccccccc}  
$<x>$ & $Q^2$ & $g_1^p/F_1^p$ & $g_1^n/F_1^n$ \\
      & (GeV$^2$) & $\pm$ stat. $\pm$ syst. & $\pm$ stat. $\pm$ syst. \\
    0.015 &  1.22 &    0.048 $\pm$    0.009 $\pm$    0.004 &   -0.125 $\pm$    0.037 $\pm$    0.006 \\
    0.025 &  1.59 &    0.057 $\pm$    0.008 $\pm$    0.006 &   -0.038 $\pm$    0.034 $\pm$    0.007 \\
    0.035 &  2.05 &    0.070 $\pm$    0.008 $\pm$    0.007 &   -0.040 $\pm$    0.034 $\pm$    0.008 \\
    0.050 &  2.58 &    0.111 $\pm$    0.009 $\pm$    0.009 &   -0.054 $\pm$    0.036 $\pm$    0.011 \\
    0.050 &  4.01 &    0.222 $\pm$    0.088 $\pm$    0.009 &   -0.852 $\pm$    0.400 $\pm$    0.012 \\
    0.080 &  3.24 &    0.155 $\pm$    0.009 $\pm$    0.013 &   -0.039 $\pm$    0.038 $\pm$    0.016 \\
    0.080 &  5.36 &    0.150 $\pm$    0.011 $\pm$    0.013 &   -0.157 $\pm$    0.048 $\pm$    0.017 \\
    0.125 &  4.03 &    0.186 $\pm$    0.012 $\pm$    0.018 &   -0.034 $\pm$    0.052 $\pm$    0.024 \\
    0.125 &  7.17 &    0.209 $\pm$    0.007 $\pm$    0.018 &   -0.066 $\pm$    0.034 $\pm$    0.025 \\
    0.125 & 10.99 &    0.307 $\pm$    0.051 $\pm$    0.018 &   -0.425 $\pm$    0.238 $\pm$    0.028 \\
    0.175 &  4.62 &    0.273 $\pm$    0.023 $\pm$    0.023 &   -0.088 $\pm$    0.106 $\pm$    0.035 \\
    0.175 &  8.90 &    0.247 $\pm$    0.012 $\pm$    0.023 &   -0.077 $\pm$    0.056 $\pm$    0.036 \\
    0.175 & 13.19 &    0.305 $\pm$    0.022 $\pm$    0.023 &   -0.010 $\pm$    0.109 $\pm$    0.039 \\
    0.250 &  5.06 &    0.358 $\pm$    0.023 $\pm$    0.030 &   -0.007 $\pm$    0.114 $\pm$    0.053 \\
    0.250 & 10.64 &    0.353 $\pm$    0.011 $\pm$    0.030 &   -0.027 $\pm$    0.056 $\pm$    0.055 \\
    0.250 & 17.21 &    0.396 $\pm$    0.014 $\pm$    0.030 &   -0.069 $\pm$    0.075 $\pm$    0.057 \\
    0.350 &  5.51 &    0.424 $\pm$    0.049 $\pm$    0.039 &    0.164 $\pm$    0.288 $\pm$    0.079 \\
    0.350 & 12.60 &    0.466 $\pm$    0.020 $\pm$    0.039 &    0.103 $\pm$    0.130 $\pm$    0.082 \\
    0.350 & 22.73 &    0.500 $\pm$    0.025 $\pm$    0.038 &    0.055 $\pm$    0.161 $\pm$    0.086 \\
    0.500 &  5.77 &    0.561 $\pm$    0.058 $\pm$    0.048 &    0.155 $\pm$    0.417 $\pm$    0.117 \\
    0.500 & 14.02 &    0.561 $\pm$    0.024 $\pm$    0.048 &    0.017 $\pm$    0.187 $\pm$    0.123 \\
    0.500 & 26.86 &    0.507 $\pm$    0.042 $\pm$    0.048 &    0.057 $\pm$    0.309 $\pm$    0.125 \\
    0.750 & 15.70 &    0.622 $\pm$    0.091 $\pm$    0.050 &    0.616 $\pm$    0.775 $\pm$    0.144 \\
    0.750 & 34.72 &    0.559 $\pm$    0.405 $\pm$    0.050 &    0.254 $\pm$    3.141 $\pm$    0.138 \\

\end{tabular}
\end{table}

\end{document}